# LARGE-SCALE STRUCTURE FROM SPATIAL CONDENSATION AND REPRODUCING DARK MASS

## Charles B. Leffert

Emeritus Professor, Wayne State University, Detroit, MI 48202
(c_leffert@wayne.edu)

**Abstract**. Accounting for large-scale structure in our universe will require not only reasonably accurate mathematical models of its cosmogony, topology and cosmology, but also a more deeper understanding of our fundamental physical concepts of *space*, *time* and *energy*. The foundations for a new "Spatial Condensation (SC-)" cosmological model for our *closed* 3-D universe has already been presented [1,2] and this paper will explore the new and different predictions for the evolution of its large-scale structure. Accounting begins with a spatial condensation solution of the "cosmological constant" problem for the infamous $10^{120}$ factor for vacuum energy. The newly derived "expansion forces" demark that distance where gravity gives way to the Hubble flow and decelerates peculiar velocities towards "rest" in the comoving reference frame. The postulated new dark mass, that scales differently than matter with the expansion, provides the seeds for the condensation of matter and the formation of black holes even before the decoupling of matter and radiation. Graphics of the predicted size of large-scale structures are presented versus astronomical measurements.

## OVERVIEW

When our spatially 3-D universe came into being, it was the surface of a small-scale 4-D structure about the size of a large beach ball, and so it certainly did not contain any of the large-scale structure that we see today such as stars, galaxies, black holes and superclusters of galaxies. At the beginning that 3-D surface contained hot plasma of uniform density that was expanding very rapidly. So how did all of today's large-scale structure come into being as that initial beach-ball size universe expanded into the enormous ball 188 trillion, trillion times its initial size?

A new cosmological model such as the one presented in the first two papers [1, 2] must also account for this large-scale structure. Our astronomers with their new powerful instruments are obtaining a wealth of new data that places an enormous strain on the present big-bang model as evidenced by the recent questionable patches to hold it together. [1]

The new spatial-condensation model with its claims to have finally defined the essence of space, time and energy, results in a tight non-forgiving model with only one adjustable parameter, its age adjustable only from 12 to 16 Gy (<u>no</u> adjustable parameters if $t_0 \approx 13.5$ Gy is accepted), and many other falsifiable predictions. In this paper the new model will be tested for its ability to account for the evolution of the large-scale structure we see around us today.

However, first we must review some basics of the new SC-model because it has features unlike any other cosmological model that will be very important for the evolution of structure.

**Basics of the SC-Model** A symmetry-breaking, spontaneous condensation of the first 4-D spatial cell in the pre-existing, higher-dimensional epi-universe began the creation of our 3-D spatial universe.[3] Thereafter, all such exposed 4-D spatial cells became catalytic sites for further spatial condensation of other 4-D spatial cells.[4]

Our universe expands because catalytic spatial condensation of very small unit cells coming from the epi-universe produce four dimensional unit cells, called "planckton" (to be abbreviated as "pk"), of Planck size ($10^{-33}$ cm) which add to the surface of a growing 4-D ball (4-D core) and that surface is our expanding 3-D universe. A planckton is 21 orders of magnitude smaller than the nucleus of an atom, far smaller than even any projected future instruments could measure. Although the detailed mechanism of the proposed spatial condensation is not yet known, it must include the production of 3-D unit cells of our 3-D interface universe. This concept of a cellular space is in sharp contrast to the present big-bang model where 3-D space is a continuum that expands and has virtual particles popping into and out of existence.

First we will explore the rate of production ($dN_4/dt$) of 4-D planckton, which defines energy in units of 4-D planckton produced per second, denoted by "pks". This definition of energy as a rate does not add anything to our 3-D universe because the product of spatial condensation, a 4-D planckton, passes on through and attaches to the 4-D core. On the other hand, "energy" can be increased or decreased in our 3-D universe by a change in the rate of spatial condensation (by the epi-universe). A good example of this failure of the conservation of energy in our 3-D universe is the very slow redshift loss of radiant energy. Since $\rho_r \propto R^{-4}$ and $E_{rad}=C^2 M_r \propto R^{-1}$, then $(dE_{rad}/dt)/E_{rad} = -H$.

The new definitions of space, time and energy will be used to solve the "vacuum-energy problem" –one of the greatest problems that confront present physics –the enormous ratio of the energy density of the universe as predicted by the particle physicist to that of the astronomer, $\rho_p/\rho_c \sim 10^{122}$, where $\rho_p \sim 10^{93}$ is the Planck density and $\rho_c \sim 10^{-29}$ g cm$^{-3}$ is the critical density..

After review of the rate of spatial condensation, the expansion forces will be derived on massive objects that are not at rest in the expanding 3-D space and finally, the contributions of the new reproducing dark mass will be analyzed.

## ENERGY ≡ RATE OF SPATIAL CONDENSATION

The gravitational and Planck constants were not explicit in the development of the creation and expansion models. Only the Planck natural units of Planck length $l_p$ and Planck time $t_p$ were used. The speed of light $C=l_p/t_p$ was introduced but only on both sides of an equation. A different third constant would set the remaining three constants: $m_p$, G and h [Normally h is Planck's constant but hereafter h=h/2$\pi$]. The gravitational constant was used in the first example of Paper 1, but in the following, the Planck mass $m_p$ will be introduced instead. Then Planck's constant $h=m_p l_p^2/t_p$ will be derived from the SC-model from which G can be expressed.

Consider the question: "What is the rate of production of 3-D space within the spherical volume of the Earth?" Normally, the Earth is considered a solid floating in an orbit around the Sun, but we know that each atom itself of the Earth is some 99.99+ % space. So in this exercise the production rate in the vacuum of 3-D space, which is the exposed surface of the 4-D core, will be calculated first and then the 3-D production rate due to the mass of the Earth will be determined. This exercise brings out some outstanding differences between the view from the SC-model and that from the present physical concepts. Also during this exercise we will derive a new source for Hubble's law.

In the relativistic model spacetime is a "flexible" [5] geometric continuum [6] while in the SC-model, both the 3-D and 4-D incompressible cellular building blocks of space are being produced in a discrete cosmic time that is defined absolutely in terms of the number of 4-D spatial building blocks that have already been produced in our 3-D universe.

In the sense of present physics, *time* does not exist in the SC-model. A changing *resistance* to spatial condensation is postulated that is called "cosmic time". Hypothetically, if spatial condensation could cease, there would be no motion and no cosmic time. Its value would have increased without limit. By the way, an examination of the partial times of Paper 2 show that creation began with cosmic time changing from a value without limit ($\infty$) to Planck time $t_p$. Cosmic time cannot have value zero.

In the parametric time of present physics, it is assumed that one unit interval of time is the same as any other and with the parametric continuum of time, there is no lower limit to the interval dt. In the SC-model cosmic time and its intervals are fixed by the definition of cosmic time.

From the SC-definition of cosmic time in terms of total mass density $\rho(R)$, $t^2 \equiv t_0^2 \rho_0/\rho(R)$, dt has a lower limit of the Planck time $t_p$ and is itself a function of cosmic time, $dt = -1/2(t_0^2\rho_0/t)(\rho_0/\rho^2(R))d\rho$. An example is the change from the early radiation-dominated era where $dt = +2t_0(\rho_0/\rho_r)^{1/2}dR/R$ to the present dark mass dominated era (and future) where $dt = +t_0(\rho_0/\rho_{DM})^{1/2}dR/R$.

At the Planck level, cosmic time increases in units of the Planck time $t_p$ and it will be shown that, for the present, the model amazingly reduces to a simple equation relating the expansion of the 4-D core and our 3-D universe:

$$(dN_4/dt)t_p/N_3 \approx 1, \qquad (0)$$

which says that there is one 4-D planckton of volume $l_p^4$ added to the 4-D core every Planck second for every 3-D planckton of 3-D space where $N_3=2\pi^2(R/l_p)^3$ and $l_p$ is the Planck length. The adverb "amazingly" is used because Eq. (0) has no dependence whatsoever on the mass-energy content of our 3-D universe in contrast to the basic equations of the SC-model in Paper 1 which attribute the relative expansion rate H <u>only</u> to the total mass-energy density of radiation, matter and dark mass of our 3-D universe,

$$H^2 = (dR/dt)^2/R^2 = [1/(t_0^2\rho_{T0})](\rho_T/\rho_{T2})\rho_T \qquad (1)$$

i.e., essentially, $H^2 \propto \rho_T$ in the SC-model, as in the big-bang model.

An enormous "vacuum energy" is implied by Eq. (0) and, at the present, it gives the same expansion rate as the mass-energy equation. We must also search for an answer to this dichotomy. Were it not for this violent activity at the Planck level predicted by Eq. (0), massive objects could not even be moved in our 3-D interface universe between the "compacted" 4-D core and the "compacted" epi-space.

Concerning the new approach to the concept of time, one additional point is that an altogether new *epi-time* would be required for the details for the actual steps in the condensation of the m-D pk epi-cells to a 4-D planckton. It was already concluded in Paper 2 that the speed of communication in epi-space must be much greater than in our 3-D space, i.e., $C^+ > 10^{24}C$.

So how should one approach this exercise for calculating the rate of production of 3-D space within the volume of the Earth and for deriving the important Eq. (0) above and Eq. (10) in Paper 2 [2]? Geometry is helpful.

The volume of a 4-D ball is $V_4=N_4 l_p^4=(1/2)\pi^2 R^4$ and the volume of its 3-D surface is $V_3=N_3 l_p^3=2\pi^2 R^3$. The time derivatives of these volumes give,

$$dV_4/dt = (dN_4/dt)l_p^4 = (2\pi^2 R^3)HR = V_3 HR, \qquad (2)$$
$$dN_4/dt = (R/3l_p^4)(dV_3/dt), \qquad (3)$$
and $$(dV_3/dt)/V_3 = 3(dR/dt)/R = 3H. \qquad (4)$$

The last Eq. (4) is the total rate of production of 3-D space in any 3-D volume $V_3$ at any radius R of the 4-D core with Hubble parameter H(R). For the present value of $H_0=68.6$ km s$^{-1}$ Mpc$^{-1}$=2.22x10$^{-18}$s$^{-1}$, $(dV_3/dt)/V_3)=6.66$x$10^{-18}$ s$^{-1}$. Inverting this number gives a value of 4.75 Gy for any present 3-D volume to reproduce its 3-D volume. A similar treatment of the 4-D core gives $(dV_4/dt)/V_4=4H$ and inverted gives 3.56 Gy for any 4-D volume (at the surface of the 4-D core) to reproduce itself.

Using Eq. (4), the radial velocity $v_3$ out of the surface $A=4\pi r^2$ of any 3-D spherical volume $V_3=(4\pi/3)r^3$ gives the "velocity-distance law",

$$v_3 = 3HV_3/4\pi r^2 = rH, \qquad (5)$$

[Hereafter, "Hubble's law"] for the expansion velocity of 3-D space (not for galaxies), but in which the galaxies tend to participate. For the Earth, $r_E$~6300 km, $v_E$=1.4 cm s$^{-1}$, far too small, as we will see, to be measurable as an expansion force, but the new 3-D spatial flow might well influence the interpretation of the Michelson-Morley experiment.[7]

Substituting Eq. (4) into Eq. (3) with $V_3=N_3 l_p^3$ and $l_p=Ct_p$ gives,

$$dN_4/dt = (N_3/t_p)((dR/dt)/C), \qquad (6)$$

that gives Eq. (0) since $(dR/dt)/C = 1.005$ at the present.

The expansion of 3-D space also occurs inside planets and even inside atoms and so-called elementary particles. However, contrary to some current thought, that small objects are not subject to the expansion [6], the spatial condensation model suggests that all massive particles bound by forces can, in principle, change size even though that change is not measurable.

To get the expansion contribution of the mass inside $V_3$ requires further manipulation of the constants. In both the Einstein-de Sitter and SC-model, the expansion rate squared in terms of the scale factor R, is proportional to the total mass-energy density as shown for the SC-model in Eq. (1), that is, the expansion rate $(dR/dt)^2$ is proportional to something residing inside the space of our 3-D universe. But the SC-model also predicts that at the present $dR/dt \approx C$ and in Eq. (6) the spatial condensation rate $dN_4/dt$ depends, not on something residing in our 3-D space, but upon $N_3$, the number itself of the total unit cells in our 3-D space. Thus there could be a density of something causing the $dN_4/dt$ <u>only</u> if it occupied every element of 3-D space.

To discover what the SC-model is telling us, expand Eq, (4), using Eq. (3), until $dN_4/dt$ is forced to contain a production rate proportional to the mass energy content of our universe and then interpret its dimensionless pre-factors.

For volume $V_3$ containing mass $M_3$, the average density is $<\rho>$ and so $V_3=M_3/<\rho>$. Now introduce the Planck density $\rho_p=m_p/l_p^3$ in terms of the Planck mass $m_p$. Rewrite Eq. (4) as,

$$dV_3/dt = V_3(3/R)l_p^4((dR/dt)/C)C^2(1/Cl_p^4). \qquad (7)$$

Since $(1/Cl_p^4)=\rho_p/h$, and $V_3$ can be replaced by $M_3/<\rho>$, one has,

$$dV_3/dt = (3/R)l_p^4(M_3C^2/h)((dR/dt)/C)(\rho_p/<\rho>). \tag{8}$$

Finally, using Eq.(3) and associating parametric time $t=\int dt$ with $(dN_4/dt)_M$,

$$(dN_4/dt)_M = M_3C^2/h, \tag{9}$$

becomes the contribution of the mass-energy $M_3C^2$ to the expansion, to give,

$$dN_4/dt = F_N(dN_4/dt)_M, \tag{10}$$

where the dimensionless $F_N$ is defined by,

$$F_N = ((dR/dt)/C)(\rho_p/<\rho>). \tag{11}$$

From the SC-computer model for $t_0=13.5$ Gy, $(dR/dt)/C=1.005$, so from Eqs. (9), (10) and (11), the above derivation from the SC-model implies that, although the right side of Eq. (9) has the correct units to express the 4-D pk production rate in terms of the mass-energy of the universe, using only the parametric-time, one must replace any density of mass at all unit volumes of 3-D space, by the Planck density. This amounts to the enormous factor of $F_N = (\rho_p/<\rho>)_0 = 10^{123}$.

Thus the SC-model leads to the amazing conclusion that the spatial condensation of "vacuum energy" is by far the greatest contribution to the expansion rate of our universe, equivalent to a Planck density in terms of parametric time, but it does not have mass and should not be included in an expansion model based on mass-energy. The defined cosmic time of the SC-model allows it to correctly predict the expansion rate in terms only of the total <u>mass-energy.</u> However, the geometric derivation of Eq.(6) allows the mass-energy to be ignored completely and give exactly the same expansion rate.

Present physics makes no attempt to define space, time or energy in terms of more fundamental concepts and so new forms such as inflationary space, repulsive energy, quintessence, etc. can be introduced without restraint. Present particle physics suggests an enormous vacuum energy, but without a definition of energy, that vacuum energy (with assumed mass) in the big-bang model would collapse our 3-D universe in only a few seconds after it came into being.

Here in the SC-model "energy", that exhibits "rest mass", is carefully defined as the rate of spatial condensation $dN_4/dt$ at its condensation site by "persistent columns of arriving m-D pk" and these columns of m-D pk also account for gravity as shown in Paper 2 [2]. Photons exhibit "mass equivalent of energy, $m=E/C^2$" when deflected by the Sun. Photons also have persistent columns of arriving m-D pk but their sites are not "persistent" for individual gravitational interaction. However a concentrated large mass-energy of radiation would dimple the 4-D core and gravitate globally.

In the above exercise, the SC-model just told us there is another mode of spatial condensation $dN_4/dt$, of far greater total magnitude $F_N$ that operates at every cell of 3-D space, but cannot form persistent columns of arriving m-D pk, since their condensation sites immediately disappear into the 4-D core, and so cannot exhibit mass and be measured; nor can they dimple the 4-D core and exhibit gravity.

New physical ideas, not higher mathematics, are needed to improve our understanding of the physical universe. Definitions promote understanding; correct definitions reveal truth and survive.

Radiation, matter and dark mass established persistent columns of incoming m-D pk to their condensation sites in the last act of creation and those columns manifest the attribute "mass". After creation, the newly produced c-type 4-D pk on the 3-D surface of the expanding 4-D core are also good condensation sites but they cannot form stable incoming columns of m-D pk. Thus to repeat, "vacuum-energy" does not have "mass" and so cannot be measured as mass[4]. However, the x-type 4-D pk are rejected by the 4-D core, remain on the surface to sustain columnar flow, and so do exhibit "mass".

Thus there are now "particles" with three different modes of spatial condensation: (1) with "rest mass", (2) with "mass but cannot rest" and (3) with "no mass".

The overview has given us a totally different picture of our universe with the new definitions of space, time and energy and a solution of the "cosmological constant" problem.[18] Next we derive effects that these new features might have on the evolution of structure. We start with the expansion forces produced by objects moving through the cellular 3-D space and then derive gravitational effects of growing clumps of dark mass.

## EXPANSION FORCES

In the SC-model with the 4-D core, the expansion of our universe is independent of gravity. Now the role of gravity takes center stage. In a universe in which the background matter mass (~baryons, $\Omega_B=0.031$) is only ~0.015 of the minimum required to spontaneously condense locally $\rho_{eq}$, see Eq. (22), the growing clumps of dark mass must initiate the birth of structure. Expansion forces oppose gravity but it will be possible to scale all of these effects with the expansion and make some predictions for the formation of structure.

Consider a galaxy of mass M and peculiar velocity $v_{pM}=0$ at a great distance from other galaxies which is gravitationally isolated except for a nearby probe mass m such as a small globular cluster as sketched in Fig. 3-1. For a spherically distributed mass surrounding a large void, Newton's iron sphere theorem and Birkhoff's theorem says that the center void is a "flat", gravity-free region of space [8]. Mass m is gravitationally bound to M. But as m is moved away from M, at some radius $r > r_{eq}$, m must be moving away from M in the Hubble flow with increasing velocity as the universe expands.

There must be some equilibrium distance from M where the gravitational force $F_G$ is balanced by an "expansion force $F_H$" and mass m, at rest with mass M, would remain so, at least temporarily, until the deceleration of the universe comes into play. If we divide both $F_G$ and $F_H$ by mass m, the equality at $r_{eq}$ becomes one of acceleration, $\mathbf{a}_G + \mathbf{a}_H = 0$.

In the SC-model $\kappa = Gt^2\rho_T$ is a universal constant (set to $3/32\pi$ as in the radiation-dominated Friedmann solution) [4] which dropped out of the derivation for cosmic time $t^2 = 1/((\rho_{T2}/\rho_{T1})^2 H^2)$ where $\rho_{T1}$ (=$\rho_T$) and $\rho_{T2}$ [1] are functions of the average densities in the 3-D universe, $1/2 \leq \rho_{T1}/\rho_{T2} = tH \leq 1$.[4] Associate $\rho_T$ with clump-mass M, $\rho_T = M/(4\pi r^3/3)$ and set $C_t = (1/8)(\rho_{T2}/\rho_{T1})^2$. For distant mass m at rest relative to mass M with $v_{pM}=0$, the total radial velocity of m from M is $v_T = Hr + v_{pm} = 0$ where $v_{pm}$ is the radial component of the peculiar velocity of m. Substituting these expressions back into the equation for $\kappa$ would give $a_G = a_r$ where $a_G = -GM/r^2$ and $a_r = +C_t H^2 r = -C_t H v_{pm}$. But there is yet another mode for the deceleration of the peculiar velocity as derived in the big-bang model, [9]

$$|v_2| = |v_1|[R(t_1)/R(t_2)] \tag{11}$$

This "drag", also not dissipative in the usual sense, would also bring the peculiar velocity of any massive object to rest in the comoving frame [9]. Equation (11) can be derived from $a_m = dv_p/dt = -Hv_p$, so the two are combined,

$$a_H = a_r + a_m = -(1+C_t)Hv_p, \tag{12}$$

and the total, $\quad a_T = a_G + a_H = -GM/r^2 - (1+C_t)Hv_p. \tag{13}$

Note for zero peculiar velocity $v_p=0$, only the gravity term survives.

At equilibrium $r=r_{eq}$, $v_T=0$ and $v_p=-Hr$. Solving for $r_{eq}$ gives,

$$r_{eq} = GM/((1+C_t)H^2)^{1/3}, \qquad (14)$$

where $3/2 \leq (1+C_t) \leq 9/8$ and $H(R)$ is the Hubble parameter. For a galaxy of mass $M=10^{12}$ $M_{sun}$ and $H_0=68.6$ km s$^{-1}$ Mpc$^{-1}$, the expansion force counters the gravitational force with, Eq. (5), $v_H=v_3==63.8$ km s$^{-1}$ at $r_{eq}=0.93$ Mpc whereas at the outer radius of the dark mass halo,[4] $r_h=0.12$ Mpc for $M_{DM}/M_m=8$, $v_3=8.23$ km s$^{-1}$ and the expansion force counters only 0.21% of the greater gravitational force. Much use will be made in the following of these three equations (12, 13 &14), as they are very important for determining the evolution of large-scale structure. [It is noteworthy that a similar derivation using the Friedmann equation of general relativity introduces the deceleration "q" in the denominator of Eq. (14) which is unacceptable for q→0 in the SC-model].

In the SC-model,[4] the rate of spatial condensation due to the kinetic energy $E_k=mv_p^2/2$ slowly decreases in the SC-model as $v_p$ goes to zero. Neglecting $dm/dt$, then $dE_k/dt = mv_p dv_p/dt = -2E_k(1+C_t)H$, so kinetic energy decays as,

$$(dE_k/dt)/E_k = -3H \text{ [radiation era]} = -(9/4)H \text{ [present]}. \qquad (15)$$

Note that contrary to some present physics [10], a mass at $v_p=0$ and Hubble velocity $v_H=v_3=Hr$, is at rest in the 3-D universe and does not constitute kinetic energy.

Now remove the two masses from Fig. 3-1 and consider the stretched string in gravity free space. Will it stay stretched or will it collapse into a ball? How long must it be to overcome its self-gravity? How long must it be to break in the center? For a nylon monofilament line of diameter 0.6 mm, linear mass of $3.4 \times 10^{-4}$ g cm$^{-1}$ and breaking strength of 400 Newton, a length $L_{eq}=105$ million miles would stay stretched and the breaking length was $L_B=72$ kpc [3,4] --- a long-range force, indeed.

"Potential energy" is not a 4-D pk production rate stored somehow in the 3-D universe. However, it does represent an increase in the 4-D pk production rate for a particle released on the 4-D dimple of another mass on the 4-D core. The potential energy per unit probe mass m is

$$u(r) = -GM/r + (1+C_t)H(v_T r - Hr^2/2). \qquad (16)$$

and the negative derivative $(-du/dr)$ gives Eq. (13). Thus for $v_T=0$, $u(r)$ is maximum when $r=r_{eq}$. The radius $|r_{eq}|$ of the maximum increases with greater age of the universe as H decreases and the sphere of radius $r_{eq}$ will be called the "Expansion-Limited Sphere of Attraction, ELSA". When $v_T=0$, $u(r)=-GM/r^2+(1/2)(1+C_t)v_H^2$ and as r increases the potential energy per unit mass in the rest frame of M goes to $\frac{1}{2}(1+C_t)v_H^2$.

Figure 3-2 presents curves of potential energy per unit probe mass m around a galaxy of mass $M_{m0}=5.8$ plus $M_{DM0}=58$ in units of $10^{10}$ $M_{sun}$ where probe mass m is at rest with M and $v_T=Hr_{eq}+v_p=0$ as in Fig. 3-1.

In principle, one could connect a distant mass m at $r>r_{eq}$ to M in such a partially restrained manner, that its partial participation in the Hubble flow generated work energy at M. But as also shown for general relativity [11], the increased $v_p$ and kinetic energy (see Eq. (15)) would increase the energy content of the universe.

At large $|r|$ in Fig. 3-2 with $v_T=0$, $v_p$ is towards M and the expansion force in the opposite direction is great to accelerate mass m into the Hubble flow. For this galaxy of $M_T=63.8 \times 10^{10} M_{sun}$, the expansion forces become negligible for the local conditions of $|r|<0.5$ Mpc where the conservation laws are in good approximation. However, for the decoupling era of Z~1000, that "local region" shrinks to $|r|<0.5$ kpc.

The expansion forces and a reproducing dark mass change the dynamics of motion significantly. An excellent example is the concept of escape velocity. For our present local physics $v_{esc}=(2GM/r)^{1/2}$ which is the radial velocity at r needed to bring v to zero as r increases without limit. This expression may be adequate to propel a mass out of the solar system but it is not adequate to propel a mass away from a source galaxy into the Hubble flow. The launch velocity for escape is a peculiar velocity. The expansion force will oppose it whether it is less than or greater than the Hubble flow $v_H=Hr$.

To obtain a measure of the new effects, the predictions of the SC-computer model for one-dimensional trajectories of a probe mass m are shown in Fig. 3-3. For the standard galaxy of $M_0=63.8 \times 10^{10} M_{sun}$, mass m was launched from the halo edge at $r_i=84.4$ kpc at three different radial velocities $v_i$.

At the standard escape velocity of $v_{esc}=235$ km s$^{-1}$, mass m returned to the galaxy in about 22 Gy and went into oscillatory motion through the galaxy. Even at launch velocity 260 km s$^{-1}$, mass m returned at ~33 Gy and went into oscillatory motion. Not until mass m was launched at 400 km s$^{-1}$ was it finally incorporated into the Hubble flow. The top curve shows that the expansion forces alone are sufficient to accelerate mass m into the Hubble flow if mass m is released at rest relative to M at a distance of r=3.9 Mpc.

## DARK MASS

Dark mass with its new scaling is key to an expansion model that accounts for the observed values of the overall cosmological parameters of our universe, e.g., its age ($t_0=13.5$ Gy), its size ($R_0=4388$ Mpc), the Hubble constant ($H_0=68.6$ km s$^{-1}$ Mpc$^{-1}$), its deceleration ($q_0=0.0082$), its mass-energy contents ($\Omega_0=0.278$) and its near steady-state expansion rate ($(dR/dt)/C=1.005$). But dark mass is also key to accounting for the major observed features of the <u>structure</u> of the matter in the present universe such as galaxies themselves, current galactic clusters, voids, walls, cold gas clouds and black holes. What detail of present structure can we hope to account for?

In our universe with the above postulated expansion forces and low mass-energy content, no structure, including stars, could have formed without the ever-growing clumps of dark mass. As opposed to the current model, with the exception of the immediate region around dark mass seeds, no matter mass could have gravitationally clumped until the decoupling of radiation from matter (Z~1000). So the 10$^{-5}$ structure in the CBR is not gravitational bound but is essentially the expanding initial distribution of dark mass seeds (some seed mergers allowed).

The voids, and the nascent hydrogen/helium gas therein, expand freely in the Hubble flow. So also does the expansion forces allow large clouds of cold hydrogen/helium, essentially free of dark mass, to remain relatively stable against either expansion or gravitational clumping. Large seeds of dark mass could form black holes even before decoupling and lead to enormous black holes in the centers of later large galaxies and exceptionally large seeds could lead to superclusters of galaxies.

In the SC-model, the creation of our 3-D universe begins with the symmetry-breaking production of the first Planck size, x-type 4-D pk of dark mass. Two types of 4-D cells are produced – a c-type and an x-type. All are produced exponentially and lead to the forced assembly of the expanding 4-D core that underlies our 3-D universe. The c-type, which are the predominant ones, are "core acceptable" and promptly loose their catalytic reproductive power as they attach to the 4-D core and are covered by the next layer of 4-D pk. The x-type are fewer because they also produce c-type 4-D pk and reproduce another x-type pk only randomly. The x-type 4-D pk are rejected flotsam on the core surface and so become *dark mass*.

The x-type 4-D pk in the clumps of dark mass continue to reproduce on the surface of the 4-D core. The arriving m-D pk from epi-space cause the dark mass to dimple the 4-D core (curve 3-D space) as explained in Paper 2 [2]. The dark mass does not otherwise interact with matter. Since both types of 4-D pk continue to reproduce on the surface, dark mass must scale with the expansion differently than matter.

During the expansion of the 3-D surface of the 4-D core of radius $R_U$, the density of radiation mass-energy scales as $R_U^{-4}$ and matter as $R_U^{-3}$, the same scale factors as for the big-bang model. However for the SC-model, the average density of dark mass must scale as $R_U^{-2}$ in order for the mass of dark mass to increase and yet have its average density decrease with the expansion, i.e.,

$$\rho_{xU} = \rho_{xU0}(R_{U0}/R_U)^2, \qquad (17)$$

where the 0-subscript refers to present values. Since the volume of 3-D space scales as $R_U^{+3}$, the mass of dark mass increases as $M_{xU}(R) = 2\pi^2 \rho_{xU0} R_{U0}^2 R_U = M_{xU0}(R_U/R_{U0}) = M_{xU0}/(1+Z)$. The time derivative shows dark mass reproduces in place at the rate,

$$dM_{xU}/dt = M_{xU}H, \qquad (18)$$

where H is the Hubble parameter.

In the absence of reasons to the contrary, a clump of dark mass should reproduce at the same rate as the total, or with local radius r, the same as with the scale factor $R_U$ or,

$$\rho_x(r) = \rho_{x0}(r_0/r)^2 \text{ and } M_x(r) = M_{x0}(r/r_0) = M_{x0}/(1+Z), \qquad (19)$$

and $\qquad dM_x/dt = M_x H. \qquad (20)$

Astronomers have found this is the way dark mass tends to scale with radius r in the halo of spiral galaxies outside the optical radius.[13]

These attributes of dark mass, "reproducing" and "in place", are very important because the combination requires the initial condition of very small seeds of dark mass at creation of the 4-D core that grow into ever larger clumps of dark mass to the exclusion of any dark mass ever being produced in the space outside those growing clumps.

Other important consequences of these two attributes are:

(a) The last "corelets" to impact and complete the formation of the 4-D core tend, like comets impacting the moon, to expel from the impact zone any already-existing surface dark mass seeds and concentrate them smoothly into a 3-D sphere surrounding a void. Any radiation or matter so concentrated would quickly disperse at the extreme temperature of creation.

(b) The predicted end-of-creation expansion rate $(dR/dt)_{eoc} > 10^{24}C$, [2] implies very important characteristics of the epi-universe: (1) that the equivalent limiting speed of epi-energy transfer is $C^+ > 10^{24}C$ which has great importance for spatial-condensation accounting of 3-D quantum behavior, and (2) because of the release of the latent epi-energy of the impacting 4-D corelets in the last act of creation could produce a surrounding bath of enormous epi-temperature to quickly homogenize the entire nascent 3-D universe of radiation and matter.

(c) The density distribution of Fig. 3-4 for large dark mass seeds, implies very early formation of dark-mass black holes which could then feed on the hot radiation-matter plasma even before decoupling of radiation and matter. As shown for the 4-D geometry of a black hole [2], the epi-universe also enters the black hole to continue spatial-condensation on the captured mass-energy and to continue reproduction of the captured dark mass that would override the Hawking evaporation.

**Dark-Mass Black Holes** The predicted dark mass density distribution is shown in Fig. 3-4 for both the standard galaxy of $5.8 \times 10^{10}$ $M_{sun}$ matter mass plus $58.0 \times 10^{10}$ $M_{sun}$ dark mass and its proto-galaxy at Z=1000 of $5.8 \times 10^{10}$ $M_{sun}$ when its dark mass was only $0.058 \times 10^{10}$ $M_{sun}$. Both distributions had the same shape with a cusp but that doesn't show on the scale of the abscissa. The bottom horizontal line shows the present average dark mass density in the universe $2.19 \times 10^{-30}$ g cm$^{-3}$ and the upper horizontal line the early average density at Z=1000 of $2.19 \times 10^{-24}$ g cm$^{-3}$.

It was also postulated that if the center region of a dark mass clump disappears into a black hole, the remaining part of the spherical clump could take up in compression the support against further collapse,[3] particularly, if captured matter mass had replaced it.

In order to develop further this notion of dark-mass black-hole formation before decoupling of radiation and matter, use will be made of Eq. (14) developed in the above section on expansion forces. This equilibrium equation also applies to mass M of a black hole and since dark mass scales as $M=M_0/(1+Z)$, these two simple equations can be used in the SC-computer model to explore the possibility for early dark mass black holes. The lower limit of mass for a sun to form a black hole is ~3 $M_{sun}$.[14] It will be assumed that this limit was also true in the past and it will be used next to predict when early dark mass black holes could have formed.

The results of the study are presented in Fig. 3-5 where the logarithm of the radius of the ELSA-sphere (log $r_{eq}$) is plotted versus the logarithm of the assumed dark mass of the black hole in units of the mass of our Sun. Start with the top curve labeled "present Z=0". Astronomers report black holes at the center of some galaxies as massive as "billions of suns". Suppose one of these monsters contained the equivalent of one billion suns ($10^9$ $M_{sun}$) of dark mass (neglect the additional matter mass) and we ask: "How did it evolve?"

Some have proposed early mergers of proto-galaxies caused large black holes. However much higher early H in Eq. (14) decreases $r_{eq}$ and thus opposes such mergers. So instead we assume no merger for this study and follow the "$10^9$-curve" back into the early universe and read the ordinate value of the dark mass contribution to the ELSA radius for capturing the surrounding matter mass to a proto-galaxy. Of course, the captured matter mass would have made the ELSA-radius even larger.

When we get back in time to the decoupling of radiation and matter at (Z+1)=1000, the black hole still contained $10^6$ $M_{sun}$ dark mass and an ELSA-radius of only 16 pc. Even earlier at (Z+1)=10,000, the dark mass was $10^5$ $M_{sun}$ i.e., much greater than the lower limit of 3 $M_{sun}$. Even if we had started with a black hole of $10^6$ $M_{sun}$, its "$10^6/(Z+1)$-curve" would pass through the point [$10^3$ $M_{sun}$] on the curve "(Z+1)=1000". So we conclude that the SC-model predicts that dark mass could have formed black holes even before the decoupling of radiation and matter.

Note that the foregoing arguments lose strength if we start with present black holes with $10^3$ $M_{sun}$ dark mass because at (Z+1)=1000 the dark mass is less than the lower limit of ~3 $M_{sun}$. Black holes in those proto-galaxies would have occurred later for Z<1000. The conclusion is that present black holes of dark mass greater than $10^3$ $M_{sun}$, could have formed well before the decoupling of matter and radiation and their expanding ELSA-spheres could have concentrated matter mass in dense "bulges" around the centers of galaxies.

**Dark-Mass Halos in Spiral Galaxies** Borgani, et al. (1999) [15], selected 58 spiral galaxies of various sizes and studied the slope of the rotational curves at the edge of the optical radius $R_{opt}$. They estimated the mass of matter in the disk $M_d$ and assumed a ratio of the mass of dark-matter to matter mass of 10 and thus correlated their data in terms of the average dark matter density versus $R_{opt}$. Using their published data, this author developed a dark mass model for spiral galaxies,[3] based on the dark-mass distribution of Eq. (19), and was able to predict the halo radius from the average dark mass density. The correlation of the data will be presented later in Fig.3-8 for the overall correlation of large-scale structure.

**Galactic Accretion of Matter Mass** Continuing with the notion that the growing seed of dark mass leads to the accretion of matter mass into the galaxy against the expansion forces, there are two opposing effects from the expansion itself, that may produce an optimum redshift Z for the accretion of matter mass.

In addition to the growing mass of dark mass with the expansion, the accretion of matter mass is aided by the decreasing Hubble parameter H which increases the ELSA-sphere for gravitational attraction. On the other hand, the background density of gaseous matter mass decreases as the inverse third power of the scale factor R.

Consider a present galaxy of total mass $M_T = M_m + M_{DM}$ where $M_m$ is the matter mass and dark mass $M_{DM} = \zeta M_m$. Neglecting collisions and mergers, consider its accreting state at past redshift Z where its mass, scaled to Z, is gravitationally bound but its matter mass is distributed uniformly out to the ELSA radius $r_{eq}$. Let the total mass at Z, $M_T' = OD(M_{mb} + M_{DMb})$ where OD, the overdensity, is the factor that $M_T'$ is greater than the background mass. The background matter mass density scales as $\rho_m = \rho_{m0}(1+Z)^3$ and $M_{mb} = (4\pi/3)\rho_{m0}(1+Z)^3 r_{eq}^3$ and since dark mass scales as $\rho_{DM} = \rho_{DM0}/(1+Z)^2$, then $M_{DMb} = M_{mb}(\zeta/(1+Z))$.

Substituting these expressions into Eq. (14) for $r_{eq}^3$, the $r_{eq}^3$ cancel and solving for the overdensity OD gives,

$$OD = 3(1+C_t)H^2/[4\pi G\rho_{m0}(1+Z)^3(1 + \zeta/(1+Z))]. \qquad (21)$$

Thus, a smaller H, a larger $\rho_{m0}$, and a larger $\zeta$, all increase the accretion of matter mass as reflected by a smaller OD inside the ELSA sphere of the proto-galaxy.

Values of overdensity versus redshift Z are shown in Fig. 3-6 for the values of the present average ratio of dark to matter mass $\zeta=5$, 10 and 20. For $\zeta=10$, the optimum redshift $Z_f$ for galaxy accretion of matter mass is $OD_{opt} \sim 45$ in fair agreement with Peebles (1993),[8] $Z_f = 40\Omega^{-1/3} = 61$ for $\Omega=0.278$.

**Other Large-Scale Massive Objects** For a summary of the predictions of the SC-model for the entire size-range of large-scale structure, there are many other structures to consider. An important goal is to show that the new concepts of a reproducing dark mass (always condensed) and the expansion forces (even inside a galaxy) are essential for explaining the evolution of large-scale structure. For example, one of the mysteries for present models is the stability of large clouds of dilute gas, i.e., how could they have survived so long against gravitational collapse?

For the predictions of the SC-model, we turn again to the gravity limiting ELSA sphere of radius $r_{eq}$ of Eq. (14) only now we ask for the average mass-density within the sphere which defines the minimum to be gravitationally bound. The ELSA density is,
$$\rho_{eq} = M/(4\pi r_{eq}^3/3) = (1+C_t)3H^2/4\pi G = 2(1+C_t)\rho_c, \qquad (22)$$
where $\rho_c$ is the critical density $\rho_c = 3H^2/8\pi G$.

Within a gas cloud of uniform density $\rho_r$ (no clumps of dark matter and neglecting pressure), consider a sphere of arbitrary radius r and mass $m=4\pi\rho_r r^3/3$. Next calculate $r_{eq}$ for mass m. In terms of radius, if $r_{eq}$ is less than r, the gas cloud is not gravitationally bound. Since $\rho_{eq} \propto r_{eq}^{-3}$, in terms of the density $\rho_{eq}$ of Eq. (22), if $\rho_r/\rho_{eq}>1$, the gas cloud is gravitationally bound and if $\rho_r/\rho_{eq}<1$, then the gas cloud is not gravitationally bound. It is expanding, albeit slowly, if $\rho_r$ is only slightly smaller than $\rho_{eq}$. As H decreases with expansion, $r_{eq}$ increases, $\rho_{eq}$ decreases, $\rho_r$ becomes greater than $\rho_{eq}$ and the cloud becomes more unstable to gravitational collapse and the temperature and pressure become more important.

Now we put the dark mass clumps back into the cloud of gas and let the densities above represent the sum of matter mass and dark mass. Nevertheless, for the arbitrary sphere of radius r, the same conditions still hold providing r is greater than the $r_{eq}$ of any cluster of clumps of dark mass. Within the large cloud of radius r, gravitationally condensation can occur within the $r_{eq}$ of any one dark mass clump. Thus the SC-model predicts that stars or regions of starburst can occur in a large, slowly expanding, gas cloud that is not gravitationally bound if it also contains clumps of dark mass.

As shown in Fig. 3-7, this ratio of densities $\rho_r/\rho_{eq}$ can be extended into the early universe to get a measure of the importance of the growing dark mass clumps to the gravitational clumping of the surrounding matter mass. For these calculations we start with matter mass $M_m=NM_{sun}$ (results are independent of N, so N=1) evenly distributed over a spherical volume $V_T=M_m/\rho_{m0}$ where $\rho_{m0}=2.72 \times 10^{-31}$ g cm$^{-3}$ is the predicted present average matter density in the 3-D universe. The amount of dark mass is varied as $M_{DM}(Z,\zeta)=\zeta M_m/(1+Z)$ where $\zeta=\rho_{DM0}/\rho_{m0}$. The total density $\rho=(M_m+M_{DM})/V_T$ and the ELSA-density $\rho_{eq}=2(1+C_t)\rho_c$ from Eq. (22). The SC-model predicts RR=$\zeta$=8.0 and the bottom curve of Fig. 3-7 predicts that evenly distributed dark mass at present eight times the matter mass would reach a maximum $\rho_r/\rho_{eq}$=0.19 at Z~45 and so would expand forever -- no stars and no galaxies without clumped dark mass.

At the other extreme, if the mass of dark mass is all in clumps of various mass and our volume $V_T$ had $M_{DM}=1.35 \times 10^4 M_m$, then (clumped or not) the entire volume $V_T$ would be gravitationally bound at Z=1000 – and once bound, it remains bound (collisions aside). A dark mass clump of $M_{DM}=250M_m$ is needed to gravitationally bind $V_T$ at Z=45. The variations of densities $\rho_T(Z)$ and $\rho_{eq}(Z)$ are also shown using the right ordinate for $M_{DM}=250M_m$. Thus clumped matter requires previously clumped dark mass in the SC-model.

In the following analysis examples of clumped-matter structures of various size are considered and an attempt is made to correlate the evolution of the size with mass following the above thesis that it is the size of the initial dark-mass clumps and the changing competition between gravity and the expansion forces that control. The grid of Fig. 3-8 has been constructed, similar to Fig. 3-5.

**Other Large-Scale Massive Structures** For a summary of the predictions of the SC-model for the entire size-range of large-scale structure, there are many more structures to consider. An important goal is to show that the new concepts of a reproducing dark mass (always condensed) and the expansion forces (even inside a galaxy) are essential for explaining the evolution of all of these large-scale structures.

The relations and tools derived above will be used to show how the predictions of the SC-model on size versus mass agree with the astronomical measurements. The grid of Fig. 3-8, similar to Fig. 3-5, will be used to show the correlation. Remember, that for $\rho_r/\rho_{eq}>1$, the mass was bound in Fig. 3-5, but in Fig. 3-8 $r/r_{eq}<1$ for the mass to be bound. The grid of Fig. 3-8 presents the theoretical predictions of the size of the ELSA-sphere, i.e., $L_{eq}$ or $r_{eq}$. The astronomical sizes are from the astronomical literature.

To explain the shape of the grid, Eq. (14) is expanded as,
$$r_{eq} = (G/(1+C_t))^{1/3}((M_{m0}+M_{DM0}/(1+Z))/ (H_0^2(1+Z)^2))^{1/3}. \qquad (23)$$
Thus, essentially, $r_{eq} \propto M^{1/3}$ to account for the increase in size with mass and $r_{eq} \propto (1+Z)^{-2/3}$ to account for the increase in size due to increasing H with decreasing Z. There is also an additional increase $\propto (1+Z)^{-1/3}$ due to increasing dark mass with decreasing Z.

The size of all objects is for the present, so they are to be compared to the top theoretical curve for $r_{eq}$ labeled "Z=0". Thus, all gravitationally bound structures should be below the Z=0 curve and they are! The same constraint would obtain for bound structures in the past at Z=45 and at Z=1000.

To briefly review these structures, begin with the top dashed curve with three calculated points for un-bound mass. According to the SC-model the large void structures should be relatively free of dark mass and should contain the nascent hydrogen-helium from the very early nucleosynthesis of the light elements. So for just matter mass in those void structures scaling as $\rho_m = \rho_{m0}(1+Z)^3$,
$$r = (3/(4\pi\rho_{m0}))^{1/3}M^{1/3}/(1+Z). \qquad (24)$$
The three calculated points, symbols "o", and the dashed curve are indeed above the Z=0 curve indicating such nascent gas is not gravitationally bound and is expanding with the void in agreement with the "RR=8" curve of Fig. 3-7.

The general trend for the bound objects in the grid of Fig. 3-8 is that the lower mass objects are more highly condensed to smaller size. For galaxies, the stars, because of their rotational momentum, tend to be far apart, and size per mass is greater. However, relative to stars in galaxies, galaxies in clusters are much closer together, so the trend of the dotted curve returns to the size $\propto M^{1/3}$ relation.

For more detail on these structures, start with our Sun of mass $M_{sun}=1.989 \times 10^{33}$ g and radius $r_{sun}=6.96 \times 10^{10}$ cm [16]. Extending the "Z=0" curve back to the ordinate, the $L_{eq}$(Sun)=$0.932 \times 10^{-4}$ Mpc or log $L_{eq}$=-4, so its ELSA-sphere range of attraction is 100 pc. There are about 40 stars within 10 pc of our Sun. As derived in the new SC-source of gravity[2] any object with mass, dimples the 4-D core and those gravitational dimples of the stars extend past their $r_{eq}$ but merge from many directions in the great dimple of the Galaxy to maintain the stability of their rotational motion. Note that the condensed Sun itself has size $2r=4.51 \times 10^{-14}$ pc, far down on an extended ordinate.

Approximate number densities and size and mass of the low-mass structures in Fig. 3-8 are from Carroll & Ostlie (1996) [14] and Binney & Merrifield (1998) [17]. The low mass condensed clouds of gas are apparently fragments of near-collisions of both early and late encounters. The mass of these objects is certainly more condensed in size relative to the dashed curve for the gas of background matter but much less than the gaseous plasma of our Sun. From Fig. 3-5, one would not expect any of these structures either to contain or to have derived from a dark mass black hole.

The spiral galaxies are not clumped at one size to mass ratio; instead they show a steady growing ratio greater than the size $\propto M^{1/3}$ relation. The dark mass halo radius $r_h$ size-versus-mass ratio was derived from theory and the spiral galaxy data of Borgani, et al. (1999) [15] assuming $\zeta=10$. Using the rotation-curve data and the new local scaling for dark mass, the author developed a spiral-galaxy model predicting the halo radius from the disk mass and optical radius.[3,4] The new scaling law $\rho_x = \rho_{x0}(r_0/r)^2$ gives,

$$r_h = (4\pi\rho_{x0}r_0^2)^{-1}M_{x0}/(1+Z), \qquad (25)$$

or $r \propto M_{x0}$, and is greater than $r \propto M^{1/3}$.

The superclusters Virgo and Coma in Fig. 3-8 appear to be a compact collection of the largest galaxies. Instead, they are probably a collection of the ordinary spectrum of galaxy sizes with one or more exceptionally massive galaxies (with high $\zeta$) to have provided the early seed for the supercluster formation. For constant size-spectrum, adding galaxies would tend to make the size to total mass ratio return to $r \propto M_T^{1/3}$.

There is an interesting twist to the conclusions on the evolution of structure to be drawn from Fig. 3.8. The SC-model demands the pre-existence of growing clumps of dark mass for clumping of matter mass; but it turns out that the most highly condensed structures of matter mass (e.g., stars) are of small size with little, if any, associated dark mass. Early dark mass seeds were indeed necessary to "attract" the matter past the ELSA-sphere density, but once past that density, where gravity and radiant cooling could take over, dark mass was not needed. It is the much less condensed, large structures that required, and still require, the presence of a large dark mass.

## SUMMARY

In summary, the new definition of space in the SC-model has predicted some amazing features of our 3-D universe. The most amazing feature is that the same expansion rate can be predicted from the mass-energy content of the universe or simply from its geometry independent of its mass-energy content. The latter case led to the prediction of an enormous vacuum energy density ($\rho_p$) that does not exhibit the attribute "mass" and so cannot be measured and solves the "cosmological constant" problem.

The production rate of 3-D space led to a new derivation of Hubble's law and to a non-dissipative expansion force on objects not at rest in the comoving frame. In particular, this expansion force defines a limit, the ELSA sphere, on the gravitational force of any massive object.

The scaling law with expansion of the new dark mass makes it an excellent seed for initiating the clumping of matter and for producing early black holes. The SC-model calculations with the expansion force showed that, for our low-mass density university, no clumped matter structure could have formed without the growing clumps of dark mass. At the same time, the expansion forces do allow the expansion of voids and stability of large clumps of cold hydrogen/helium gas.

In the future, distant clumps of matter mass will become more and more isolated with cosmic time but there is much more clumping of matter mass yet to occur within the growing reaches of the ELSA spheres.

The author thanks his good friend, Emeritus Professor Robert A. Piccirelli, for extensive discussions of the new physical concepts.

## NEXT PAPER 4

Astronomers obtain their basic information from radiation received from distant structures in our universe – its direction of reception, the magnitude of its luminosity and the flux and spectrum of its arriving energy. The interpretation of this information depends strongly on the astronomer's model of how our universe is constructed and how it expands now and in the past as was made so clear by the difference of interpretation of the supernova Ia data in paper 1. [1] Interpretations of received radiation would vary greatly between an assumed closed universe and an assumed $\Omega=1$ flat universe. Radiation in a closed universe will be the topic of the next paper 4. Many new falsifiable predictions will be made.

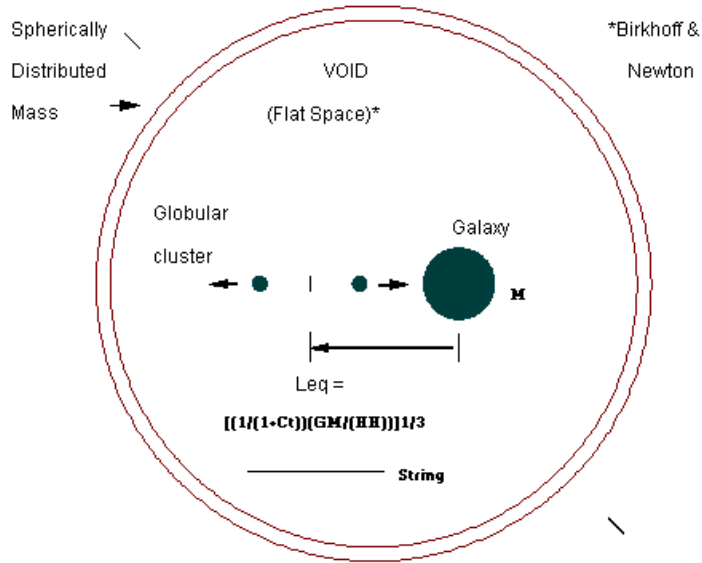

Fig.3-1. The above void within the spherically distributed mass (without M and m) would be a region free of gravity. The SC-model agrees, but it also predicts for a small probe mass m at rest with a galaxy of mass M within the void, there would be a radius $r_{eq}=L_{eq}$ where the outward expansion force would just balance the inward force of gravity.

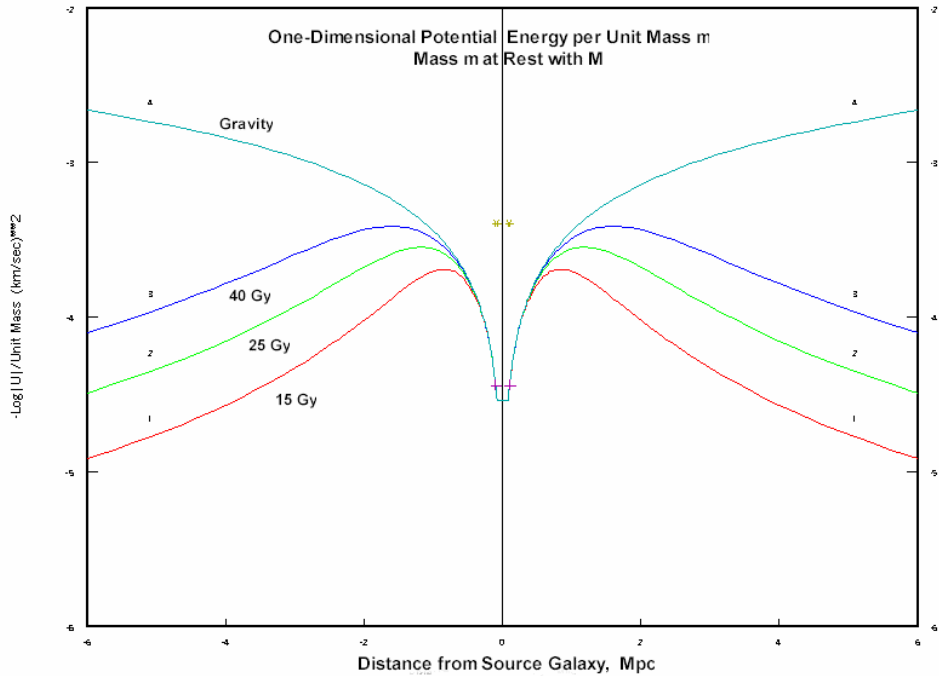

Fig. 3-2. In the SC-model, potential energy represents an increase in spatial condensation that could be obtained from epi-space, not an "energy" that already exists. Here for small mass m at rest with galaxy mass M, the negative slope of the potential energy curves show the change with time of the magnitude and direction of the net radial force between expansion and gravitation. Within ~ 0.5 Mpc of this galaxy, our standard physics is in excellent approximation. Three curves show the evolution with age due to decreasing $H_0$.

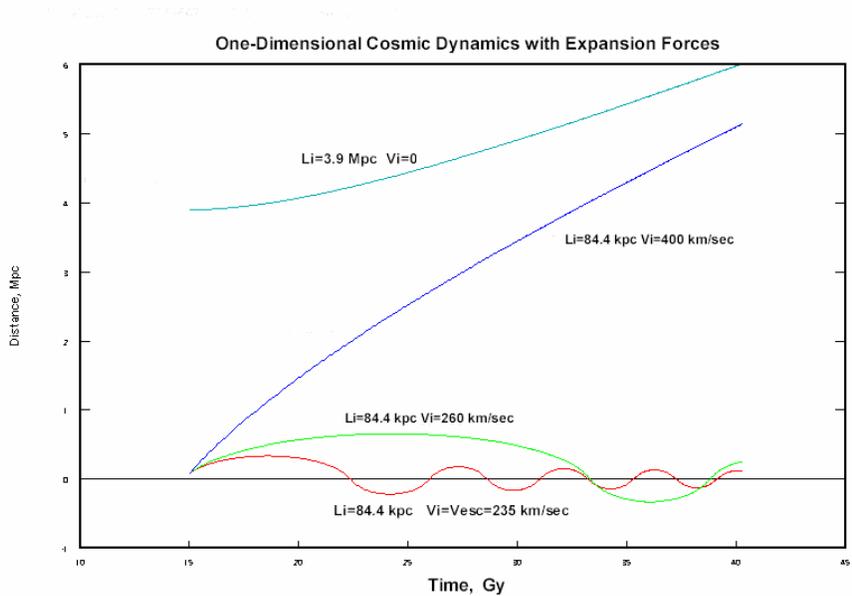

Fig. 3-3. The expansion force opposes any peculiar velocity $v_p$ and therefore requires an escape velocity greater than the standard physics expression, $v_{esc} = (2GM/r)^{1/2}$ as indicated by the lower two curves for a galaxy of $M_T=63.8\times10^{10}M_{sun}$. Not until the launch velocity was 400/235=1.70 greater, was probe mass m put safely into the Hubble flow. From a position at rest at r=2 Mpc relative to M, the expansion force was sufficient itself to propel m into the Hubble flow.

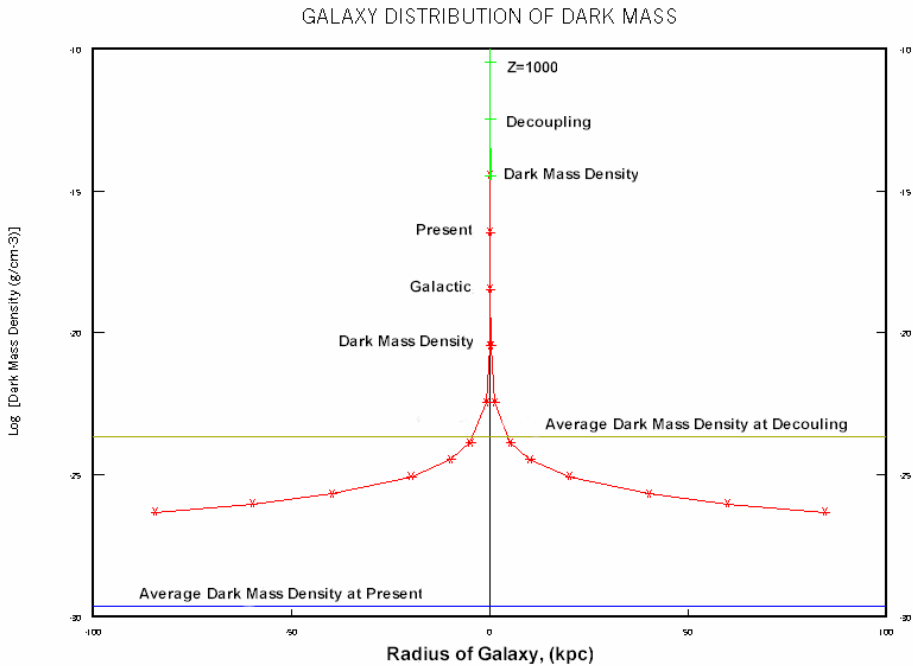

Fig. 3-4. The natural local density distribution of dark mass $\rho_x=\rho_{x0}(r_0/r)^2$ is postulated to be the same power law as the average dark mass density of the university varies with the expansion of its radius $\rho_{xu}=\rho_{xu0}(R_{u0}/R_u)^2$. The predicted center cusp could promote early dark mass black holes. The top three points represent the distribution at Z=1000.

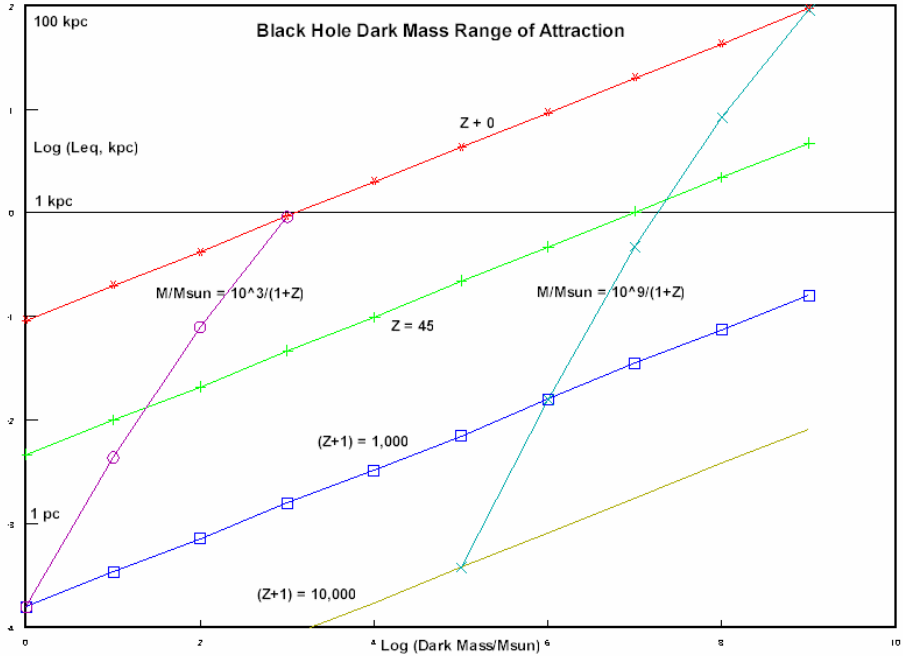

Fig. 3-5. The expansion-limited-sphere-of-attraction "ELSA" of radius $r_{eq}=L_{eq}$ for dark-mass black holes varies with mass as $M^{1/3}$ and with redshift as $(1+Z)^{-1}$. Fixing the lower limit of $M>3M_{sun}$, present galactic-centered black holes from $10^3$ to $10^9$ $M_{sun}$ could have been produced by dark mass before Z=1000.

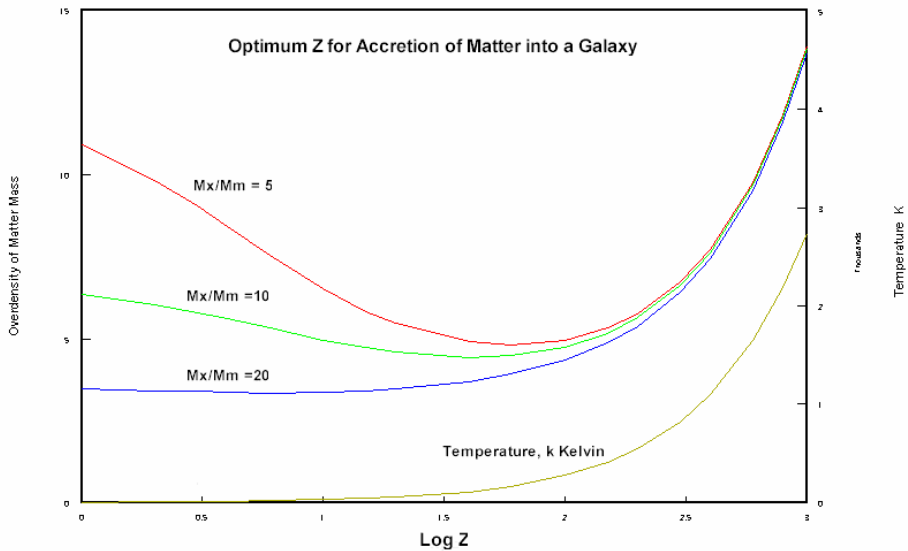

Fig. 3-6. The limiting radius $r_{eq}$ of the ELSA-sphere increases with increasing dark mass, accreted matter mass and decreasing expansion rate H. But the background matter density decreases with the expansion as $\rho_m=\rho_{m0}(R_{u0}/R_u)^3$. For optimum matter accretion into a growing proto-galaxy, all these effects combine to favor a minimum overdensity OD of matter within $r_{eq}$ at $Z_f\sim 45$ for $\zeta=(\rho_{DM}/\rho_m)_0=10$.

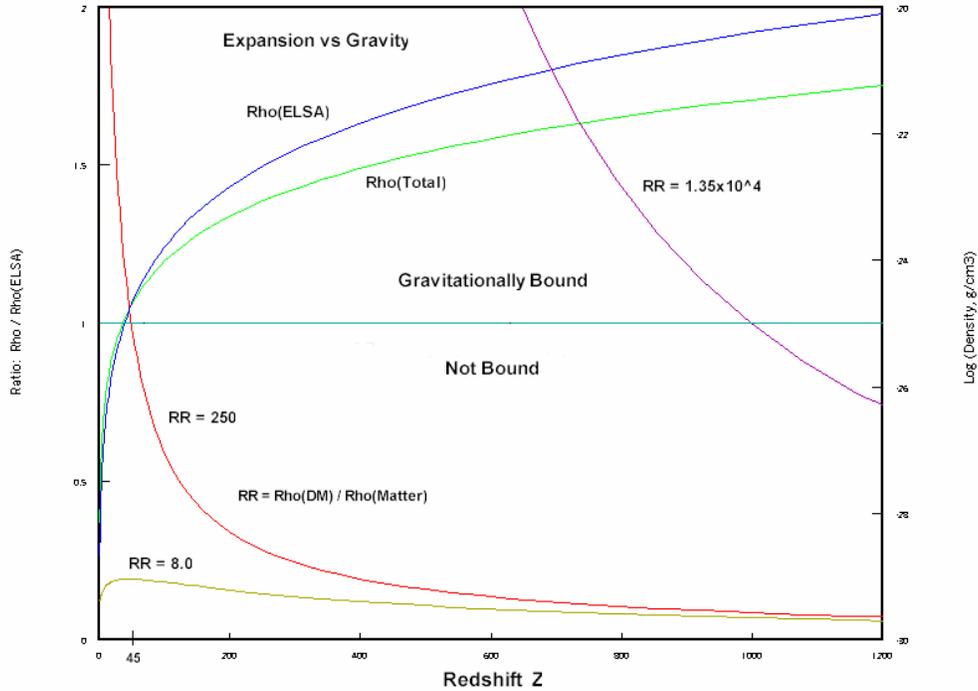

Fig. 3-7. The RR=$\rho_{DM}/\rho_m$=8 curve shows that no stars or galaxies could have formed if dark mass was evenly distributed (see text). On the other hand, the mass would have been gravitationally bound at Z=1000, if a volume had RR=$1.35\times10^4$ more dark mass clumps than matter mass or for RR=250, the mass would have been bound at Z=45.

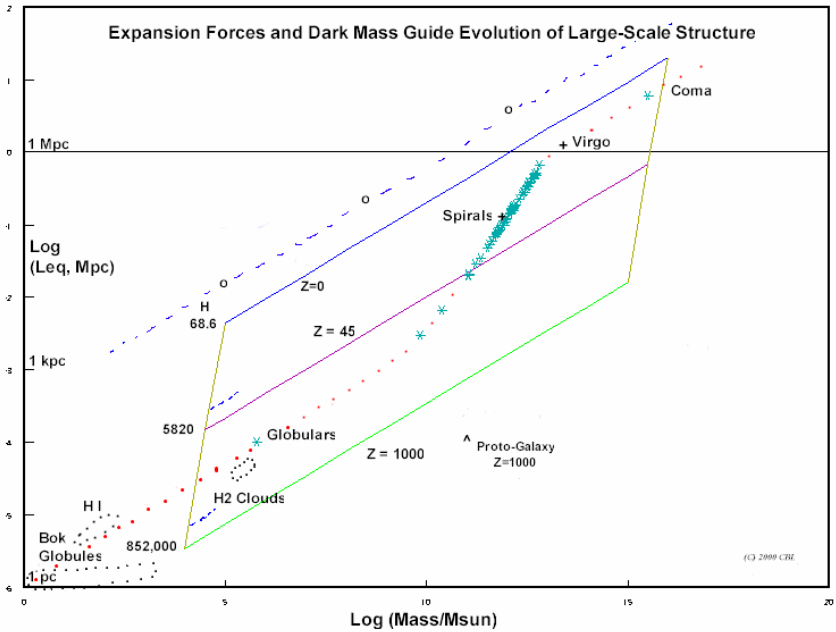

Fig. 3-8. The size versus mass dependence of selected examples of large-scale structures in our universe is overlaid with the theoretical Z-dependent grid of the ELSA-sphere radius $r_{eq}$ versus mass. All gravitationally bound structures must be below the Z=0 curve and they are. The top dashed curve represents unbound structures if they exist. Large initial seed clumps of dark mass drive the evolution of large-scale bound structures.